\documentclass[preprint2]{aastex}
\pdfoutput=1
\usepackage{spr-astr-addons}
\usepackage{graphicx}

\begin{document}
\title{Doppler Boosting, Superluminal Motion, and the Kinematics of AGN Jets}
\shorttitle{AGN Jets}
\shortauthors{Kellermann et al.}

\author{K.~I.~Kellermann}
\affil{National Radio Astronomy Observatory,
       520 Edgemont Road, Charlottesville, VA~22903--2475, U.S.A.}
\email{kkellerm@nrao.edu}

\author{Y.~Y.~Kovalev}
\affil{Max-Planck-Institut f\"ur Radioastronomie, Auf dem H\"ugel 69, 53121 Bonn, Germany, and
       Astro Space Center of Lebedev Physical Institute, Profsoyuznaya 84/32, 117997 Moscow, Russia}

\author{M.~L.~Lister}
\affil{Department of Physics, Purdue University, 525 Northwestern Avenue, West Lafayette, IN 47907, U.S.A.}

\author{D.~C.~Homan}
\affil{Department of Physics and Astronomy, Denison University, Granville, OH 43023, U.S.A.}

\author{M.~Kadler}
\affil{Astrophysics Science Division, NASA Goddard Space Flight Center, Greenbelt Road, Greenbelt, MD 20771, U.S.A.}

\author{M.~H.~Cohen}
\affil{Department of Astronomy, Mail Stop 105-24, California Institute of Technology, Pasadena, CA 91125, U.S.A.}

\author{E.~Ros, J.~A.~Zensus}
\affil{Max-Planck-Institut f\"ur Radioastronomie, Auf dem H\"ugel 69, 53121 Bonn, Germany}

\author{R.~C.~Vermeulen}
\affil{Netherlands Foundation for Research in Astronomy, Postbus 2, NL-7990 AA Dwingeloo, Netherlands}

\author{M.~F.~Aller and H.~D.~Aller}
\affil{Department of Astronomy, University of Michigan, 817 Denison Building, Ann Arbor, MI 48109--1042, U.S.A.}

\begin{abstract}
We discuss results from a decade long program to study the fine-scale
structure and the kinematics of relativistic AGN jets with the aim of
better understanding the acceleration and collimation of the relativistic
plasma forming AGN jets.  From the observed distribution of brightness
temperature, apparent velocity, flux density, time variability, and
apparent luminosity, the intrinsic properties of the jets including
Lorentz factor, luminosity, orientation, and brightness temperature are
discussed. Special attention is given to the jet in \object{M87}, which
has been studied over a wide range of wavelengths and which, due to its
proximity, is observed with excellent spatial resolution. 

Most radio jets appear quite linear, but we also observe curved non-linear
jets and non-radial motions.  Sometimes, different features in a
given jet appear to follow the same curved path but there is evidence
for ballistic trajectories as well.  The data are best fit with a
distribution of Lorentz factors extending up to $\gamma\sim 30$ and
intrinsic luminosity up to $\sim 10^{26}$\,W\,Hz$^{-1}$.  In general,
gamma-ray quasars may
have somewhat larger Lorentz factors than non gamma-ray quasars. 
Initially the observed brightness temperature near the base of the jet
extend up to $\sim 5\times 10^{13}$~K which is well in excess of the
inverse Compton limit and corresponds to a large excess of particle
energy over magnetic energy.  However, more typically, the observed
brightness temperatures are $\sim 2 \times 10^{11}$~K, i.e., closer
to equipartition.

\end{abstract}

\section{Introduction}
\label{sec:intro}

More than 40 years ago, \cite{S64} argued that the optical jet in \object{M87}
and in quasars such as 3C 273 appeared anisotropic due to differential
Doppler boosting.  Shklovsky realized that since radio galaxies such as
Cygnus A typically have symmetric lobes, the jets which feed the radio
lobes must also be two sided, but that they appear one sided due to
differential relativistic Doppler beaming.

For the past decade, we have been using the NRAO Very Long Baseline
Array (VLBA) at~2 cm wavelength to study the relativistic flow in AGN jets
\citep{2cmPaperI,2cmPaperIII,2cmPaperII,2cmPaperIV,L05,LH05}. These
observations have a nominal resolution of about 1 milliarcsec (mas) and
are complemented by multi-epoch spectral observations at the
\mbox{RATAN-600} \citep{Kovalev_etal99,KKNB02} and University of
Michigan radio telescopes \citep{Aller_etal85,AAH03} to determine
``light-curves'' over a wide range of frequencies.  Our 2~cm VLBA
observations are also complemented by observations made by \cite{VBT03} who have used the VLBA at
6~cm to study motions in nearly 300 sources from the Caltech-Jodrell
Flat Spectrum sample. \cite{J01,J05} have observed a smaller number of
sources at multiple epochs with the VLBA at 7mm wavelength with even
higher angular resolution, while \cite{P07} have used the VLBA along
with a global array to study jet kinematics at 3.5~cm.

It is widely believed that the central engine which powers these radio
and optical jets is due to accretion onto a super-massive black hole.  We
want to know where and how the jet plasma flow gets collimated and accelerated to
relativistic velocities and whether or not there are accelerations or
decelerations along the jet.  We also want to know the intrinsic Lorentz
factor, $\gamma=(1-\beta^{2})^{-1/2}$, the intrinsic luminosity, $L_o$,
and the intrinsic brightness temperature, $T_o$, where
$\beta$ is the speed of the
relativistic plasma normalized to the speed of light.
What determines these
intrinsic jet properties and are they are related to other observables
such as the variability or luminosity at other wavelengths?

\subsection{Basic Relations}
\label{sec:equations}

All quantitative values given in this paper are based on a cosmology
with $H_0 = 70$~km\,s$^{-1}$\,Mpc$^{-1}$, $\Omega_\mathrm{m}=0.3$, and
$\Omega_\Lambda=0.7$.

Due to relativistic effects, we observe apparent jet speeds,
luminosities, and brightness temperatures which are related to the
corresponding intrinsic quantities in the AGN rest frame
through the Doppler factor, $\delta$, the Lorentz factor, $\gamma$, and
the jet orientation, $\theta$, with respect to the line of sight.

The apparent velocity, $\beta_\mathrm{app}$, the apparent
luminosity, $L$, the apparent brightness temperature, $T_\mathrm{app}$
and the Doppler factor, $\delta$, can be calculated from the Lorentz
factor, $\gamma$, the angle $\theta$ to the line of sight, and the
intrinsic luminosity, $L_o$.

The apparent velocity $\beta_\mathrm{app}$ is given by
\begin{equation}
\beta_\mathrm{app} = \frac{\beta\sin\theta}{1-\beta\cos\theta}\,,
\label{eq:beta_app}
\end{equation}
and the apparent luminosity, $L$, by
\begin{equation}
L = L_o \delta^n\,,
\label{eq:lum}
\end{equation} 
where the Doppler factor, $\delta$, is  
\begin{equation}
\delta = \gamma^{-1}(1-\beta\cos\theta)^{-1}\,,
\label{eq:delta}
\end{equation}
and where
$L_o$ is the luminosity that would be measured by an observer in the AGN
frame, and $n$ depends on the geometry and spectral index and is
typically in the range between 2 and 3. 
The apparent brightness temperature, $T_\mathrm{obs}$ is given by
\begin{equation}
T_\mathrm{obs} = \delta T_\mathrm{int}\,,
\label{eq:Tb}
\end{equation}
and the Lorentz factor, $\gamma$ by
\begin{equation}
\gamma = {(1-\beta^2)}^{-{1/2}}.
\label{eq:lorentz}
\end{equation}

\begin{figure}[t]
\begin{center}
\resizebox{1.0\hsize}{!}{
   \includegraphics[trim= 9cm 14cm 1.2cm 6.6cm]{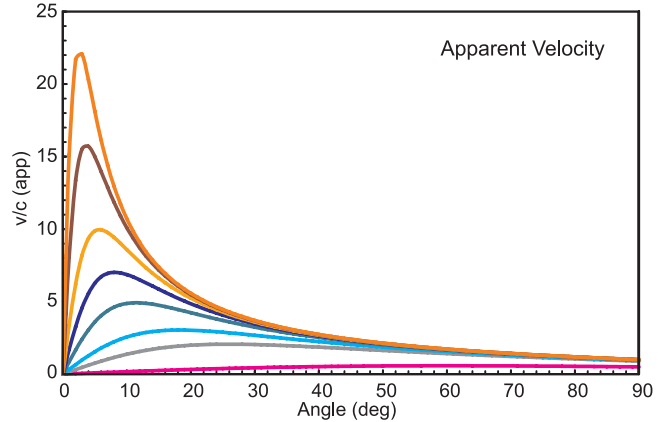}
}
\end{center}
\caption{\label{beta_app}
Apparent velocity vs.\ orientation for various values of intrinsic
velocity.  Red: $\beta=0.5$, $\gamma=1.15$; grey: $\beta=0.9$,
$\gamma=2.3$; light blue: $\beta=0.95$, $\gamma=3.2$; green:
$\beta=0.98$, $\gamma=5.0$; purple: $\beta=0.99$, $\gamma=7.1$; yellow:
$\beta=0.995$, $\gamma=10.0$; brown: $\beta=0.998$, $\gamma=15.8$;
orange: $\beta=0.999$, $\gamma=22.4$.
}
\end{figure}

\begin{figure}[t]
\begin{center}
\resizebox{0.5\hsize}{!}{
   \includegraphics[trim= 7.8cm 8.0cm 7.8cm 3.8cm]{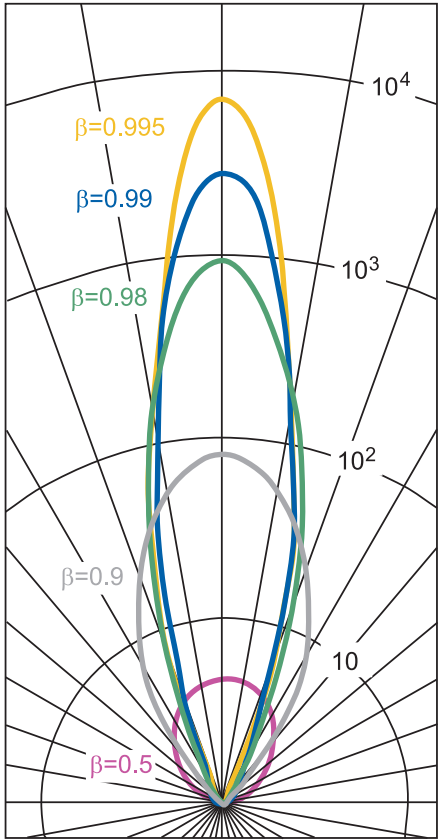}
}
\end{center}
\caption{\label{luminosity}
Luminosity Doppler boosting factor for the case where $n$=3 shown in
polar coordinates.  The radial lines indicate angles at intervals of 10
degrees and the circles the luminosity boosting factor. Red:
$\beta=0.5$, $\gamma=1.15$; grey: $\beta=0.9$, $\gamma=2.3$; green:
$\beta=0.95$, $\gamma=3.2$; blue: $\beta=0.98$, $\gamma=5.0$;
$\beta=0.99$, $\gamma=7.1$; $\beta=0.995$, $\gamma=10.0$.
}
\end{figure}

In Figure~\ref{beta_app}, we show a plot of the apparent velocity,
$\beta_\mathrm{app}$, versus orientation angle for various values of
intrinsic speed, $\beta$.  The maximum velocity is equal to $\gamma$ and
occurs at an angle $\theta_\mathrm{c} = \gamma^{-1}$. When $\theta =
\theta_\mathrm{c}$, $\delta\sim\beta_\mathrm{app}$.
Figure~\ref{luminosity} shows the effect of Doppler boosting which
reaches  a factor of $\sim 10^4$ for $\gamma=10$ ($\beta=0.995$) and
$n=3$.


Assuming that the observed jet speeds, which are due to pattern motion,
reflect the bulk jet flow velocity which is responsible for luminosity
beaming, equations \ref{eq:beta_app} to \ref {eq:lorentz} may be used to
derive the intrinsic values of $\gamma$, $\delta$ and $T_\mathrm{b}$.
 
Due to inverse Compton cooling, the maximum sustained peak intrinsic
brightness temperature, $T_\mathrm{int}$, is less than about
$10^{11.5}$~K.  Close to this value, the energy in relativistic
particles greatly exceeds the energy in magnetic fields \citep{KPT69}. 
If on the other hand the source is near equilibrium where the particle
energy is close to the magnetic energy, then, $T_\mathrm{int}\sim
10^{10.5}$~K \citep{R94}.

\section{Radio jet structure}
\label{sec:structure}

\begin{figure}[t]
\begin{center}
\resizebox{0.75\hsize}{!}{
   \includegraphics[trim = 5cm 7cm 5cm 6.2cm]{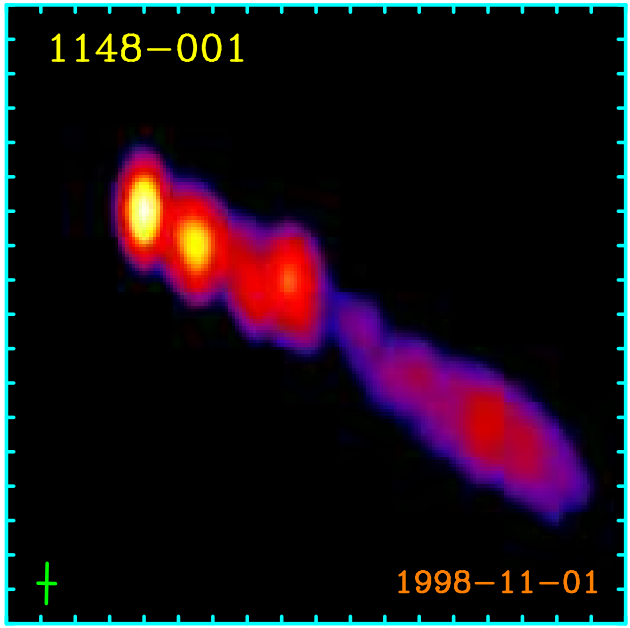}
}
\end{center}
\caption{\label{1148}
2~cm VLBA image of the one sided jet in the quasar PKS~1148$-$001.  The tick marks are spaced 
1 mas apart.}
\end{figure}

\begin{figure}[h!]
\begin{center}
\resizebox{0.75\hsize}{!}{
   \includegraphics[trim = 5cm 7cm 5cm 6.2cm]{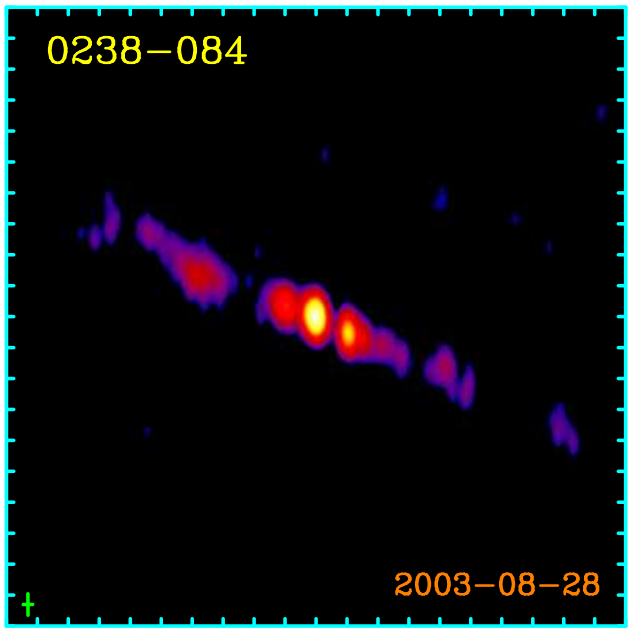}
}
\end{center}
\caption{\label{NGC1052}
2~cm VLBA image of the symmetric jets in the radio galaxy NGC~1052.  The tick marks are 
spaced 2 mas apart.  
}
\end{figure}

As a result of Doppler boosting, most AGN jets such as seen in 
\object{PKS~1148$-$001} (Figure~\ref{1148}) appear asymmetric. However
some jets with small Doppler factors, especially those associated with
galaxies such as \object{NGC~1052} (Figure~\ref{NGC1052}) appear more
symmetric.

\begin{figure}[t]
\begin{center}
\resizebox{0.75\hsize}{!}{
   \includegraphics[trim = 5cm 7cm 5cm 6.2cm]{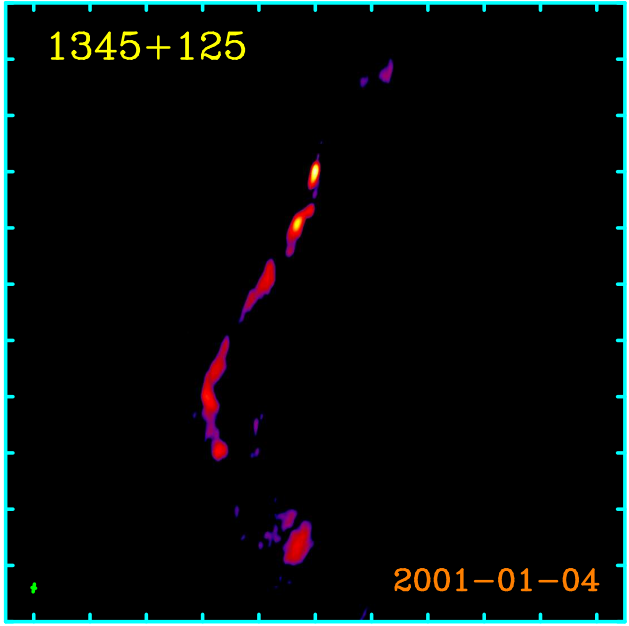}
}
\end{center}
\caption{\label{1345}
2~cm VLBA image of the curved jet in the GPS radio galaxy PKS~1345+125.  The tick marks are spaced 10 mas apart.}
\end{figure}

\begin{figure}[h!]
\begin{center}
\resizebox{0.75\hsize}{!}{
   \includegraphics[trim = 2cm 1.0cm 2cm 0cm]{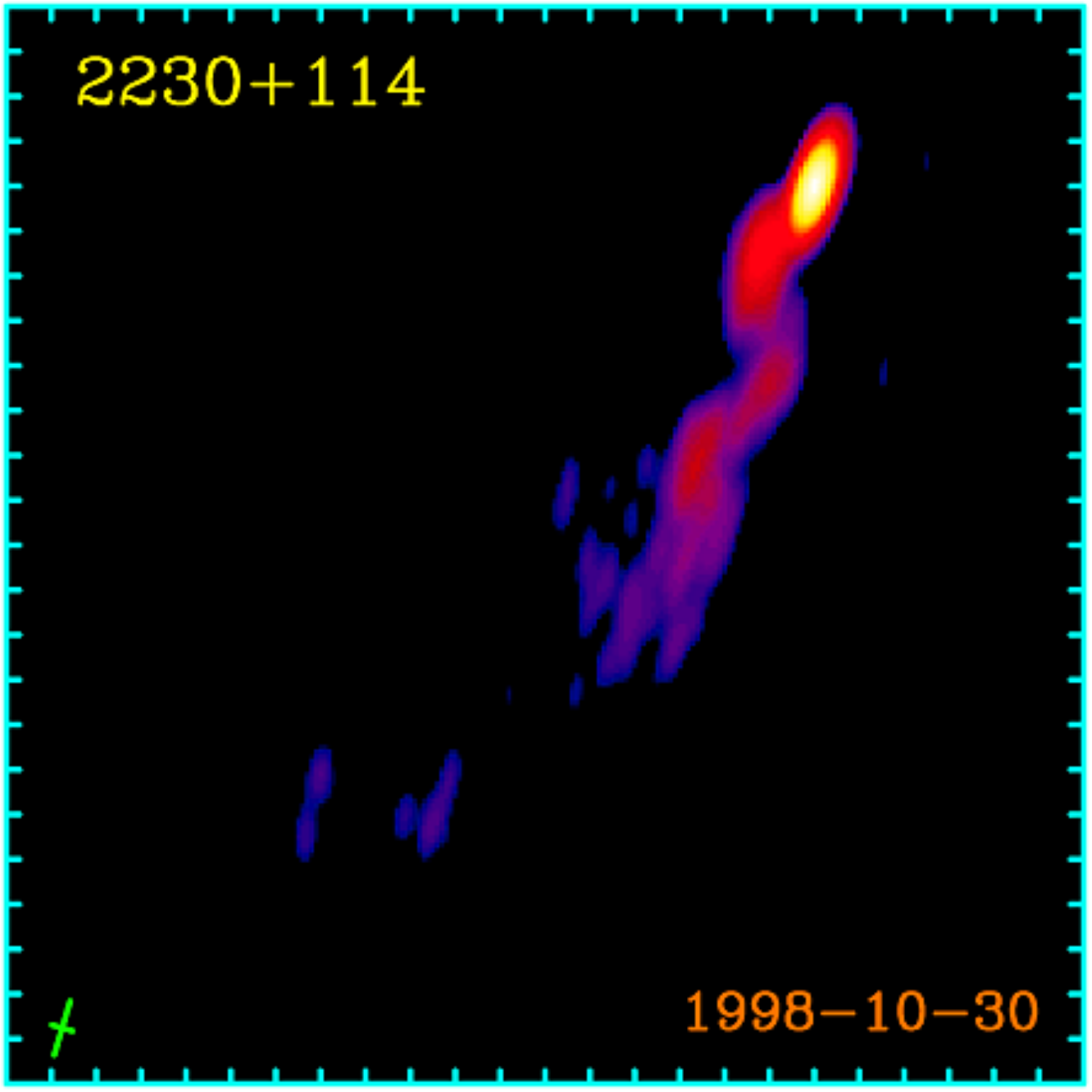}
}
\end{center}
\caption{\label{CTA102}
2~cm VLBA image of CTA 102 showing a helical jet.
The tick marks are spaced 2 mas apart.}
\end{figure}

Most of the jets we have observed are fairly straight, but some, such as
the GPS radio galaxy \object{PKS~1345+125} have pronounced curvature (see
Figure~\ref{1345} and \citealt*{LKV03}). In other sources, such as
\object{CTA~102} (Figure~\ref{CTA102}) and \object{BL~Lac}
\citep{DMM00}, there is evidence of a helical structure.


In more than half of the sources, the base of the jet appears unresolved
and is smaller than 0.05~mas in at least one dimension
\citep{2cmPaperIV}. This has interesting implications for jet physics.
(See Section~\ref{sec:beta_T}.)

\subsection{The radio galaxy M87}
\label{sec:M87}

\begin{figure}[t]
\begin{center}
\resizebox{\hsize}{!}{
   \includegraphics[trim= 0cm 0.9cm 7.2cm 20.5cm]{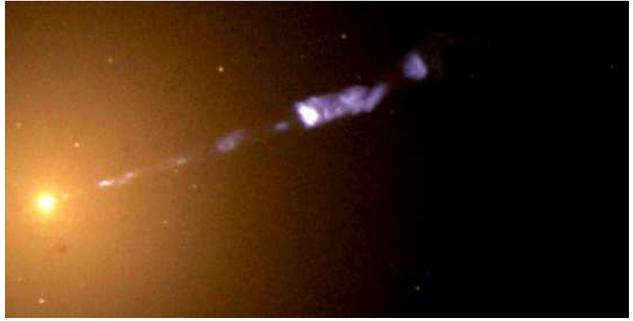}
}
\end{center}
\caption{\label{M87_HST}
HST image of the M87 jet.
}
\end{figure}
\begin{figure}[h!]
\begin{center}
\resizebox{\hsize}{!}{
   \includegraphics[trim = 0cm 0.9cm 5.7cm 19.3cm]{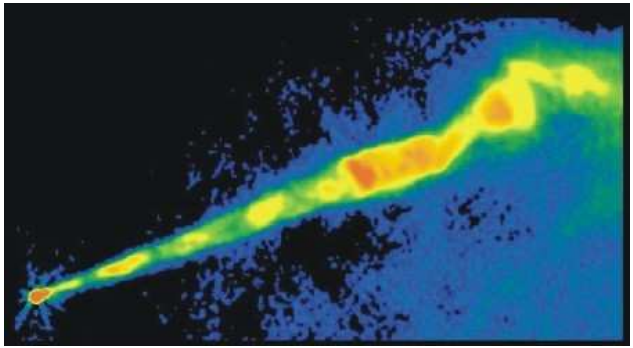}
}
\end{center}
\caption{\label{M87_VLA}
VLA 2~cm image of the M87 jet made with a resolution of 0.1 arcsec or 8~pc.
Taken from \cite{OHC89}.
}
\end{figure}
\begin{figure}[h!]
\begin{center}
\resizebox{\hsize}{!}{
   \includegraphics[trim = 3.2cm 11.2cm 3.2cm 10.3cm]{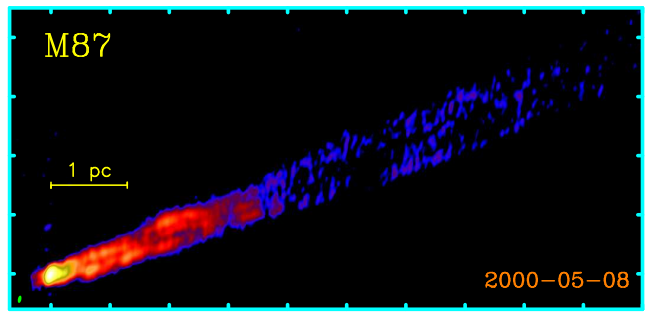}
}
\end{center}
\caption{\label{M87_VLBA}
VLBA 2~cm image of the M87 jet. The tick marks are spaced 10 mas apart, and 
the resolution is about 1 mas = 0.08
parsec.  This high dynamic range image which traces the jet out to
nearly 0.1 arc sec was obtained from a full sky track using the VLBA
along with one VLA antenna to increase the effective field of view.
Adapted from \cite{Kovalev_M87_2007}.
}
\end{figure}

The radio source associated with \object{M87} was one of the first to be
recognized as being of extragalactic origin.  It remains of great
interest today, since at a distance of 16 Mpc, it is one of the 
closest radio galaxies.  The VLBA
linear scale of 1 mas = 0.08 pc is sufficient to resolve the \object{M87} jet
transverse to its extent and to study kinematics within a few tenths of
a parsec from the central engine.

Figure \ref{M87_HST} shows the HST optical image of the \object{M87}
jet, and Figure \ref{M87_VLA} the VLA 2~cm image, each with a resolution
of about 0.1~arcsec.  The morphology at radio, optical, and even X-ray
wavelengths is very similar, suggesting a common synchrotron radiation
mechanism at all wavelengths. At least at optical and X-ray wavelengths,
the electron lifetime down the jet is much shorter than
the travel time from the nucleus, so there must be continual acceleration
of relativistic particles within the jet itself \citep{HK06}.

In Figure \ref{M87_VLBA}, we show the 2~cm VLBA image with a resolution
of about 1~mas \citep{Kovalev_M87_2007}. The VLBA image shows an
apparent bifurcation of the inner jet, starting about 5~mas (0.4~pc)
from the core.  However, it is not clear if the apparent gap which
extends along the center of the jet axis is due to an actual splitting 
of the jet, or if there is a thin cylindrical jet which appears limb
brightened.  The appearance of limb brightening may be the result of a
spine-sheath configuration  in which a faster inner jet is beamed in a
more narrow cone away from the observer.  The VLBA image also shows
evidence of an apparent ``counter-jet'' extended  toward the southeast. 

Observations of the \object{M87} jet made over more than a decade by
\cite{Kovalev_M87_2007} indicate apparent jet flow speeds
ranging from nearly stationary up to about $0.6c$.  This is
characteristic of radio galaxy jets such as observed in
\object{NGC~1052} \citep{V03}.

Considering that \object{M87} was suggested by Shklovsky as the
archetypical asymmetric Doppler boosted jet, its apparent velocity
appears surprisingly ``slow,''  although from equations \ref{eq:lum} and
\ref{eq:delta}, for orientations within about $30^\circ$ from the line of
sight, the jet to counter-jet ratio can be up to several orders of
magnitude even for mildly relativistic velocities with $\gamma~\sim{2}$. 
It should also be noted that faster speeds have been observed further 
downstream \citep[e.g.,][]{BSM99}.

\section{Jet Kinematics}
\label{sec:kinematics}

Repeated observations of quasar jets show values of $\beta_\mathrm{app}$
ranging up to about 30, corresponding to intrinsic velocities greater
than 99.9 percent of the speed of light, while radio galaxy and BL Lac jets are
typically much slower with subluminal apparent velocities.  For example, in 
NGC 1052 we find only relatively slow jet motions with a velocity of 0.27c.  We only see
outward flows and do not see any motions with significant inward motion
toward the jet base, although many features appear stationary.  While there
are significant differences in the apparent velocity of different jet
features, the velocity spread from source to source is greater than
within an individual jet. \cite{2cmPaperIII} suggested that there is a
characteristic speed for each jet which is probably reflected by the
fastest observed jet feature.  

In general, the observed motions can usually be described by a uniform
velocity which is pointed radially outward from the base of the jet. 
Sufficiently long VLBA time baselines are now being obtained so that it is
possible to detect in some individual cases non radial motions.  In some
jets the motion appears to follow the pre-existing path of earlier
features \citep{2cmPaperIII}.  In other sources, the motion appears
ballistic, but with different features moving in different directions,
and there is some evidence for periodicities which may be the result of
a precessing jet nozzle. \cite{Homan_etal03} have reported an abrupt
change in the apparent trajectory of the nearly aligned jet in
\object{3C~279} and have suggested that the final jet collimation is still occuring
more than a kiloparsec downstream from the base.

In order to understand the jet physics, we want to be able to derive the
intrinsic parameters from the observations of apparent speed, luminosity,
and brightness temperature.  We consider three approaches to estimate
the Doppler factors along with the intrinsic Lorentz factor, rest frame 
luminosity, and rest frame brightness temperatures.

\subsection{The $\beta_\mathrm{app}$ -- luminosity relation}

\begin{figure}[t!]
\begin{center}
\resizebox{\hsize}{!}{
   \includegraphics[trim = 1.1cm 6.5cm 1.2cm 3.5cm]{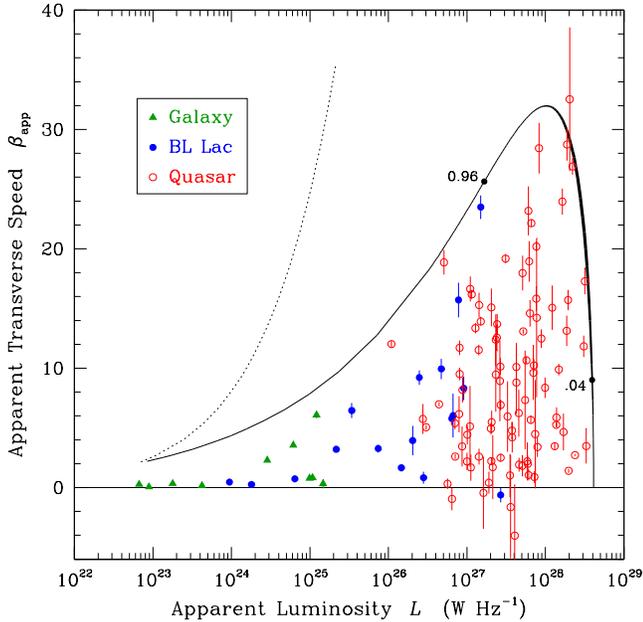}
}
\end{center}
\caption{\label{beta_lum}
Plot of apparent transverse speed, $\beta_\mathrm{app}$, vs.\ apparent
luminosity, $L$, for the fastest component in 119 well observed sources
in the 2~cm VLBA surveys. The curve shows the locus of points,
($\beta_\mathrm{app}, L$), for sources with $\gamma_\mathrm{max}\approx
32$, and $L_{o}\sim 10^{26}$\,W\,Hz$^{-1}$. The viewing angle increases
from zero going from right to left.  The thickness of the line
corresponds to the probability that a source will be observed at a given
location on the curve, and the locations of the 4 and 96 percent
probability points are labeled.  The dotted line represents an
observational limit determined by the weakest sources included,
$S_\mathrm{VLBA,med}=0.5$\,Jy, and the fastest detectable angular motions
of $\mu=4$\,mas\,yr$^{-1}$ set by our sampling interval of a few times
per year. Red open circles represent quasars; blue full circles, BL~Lacs;
and green triangles, galaxies. Adopted from \cite{C06}.
}
\end{figure}

From equations \ref{eq:beta_app} and \ref{eq:lum} both the apparent
velocity and apparent luminosity depend on the intrinsic speed and
orientation.  Thus, we might expect to see a relation between observed
speed and observed luminosity.  \cite{C06} have shown that the
distribution of observed speed and luminosity is consistent with
relativistic beaming models having a maximum value
$\gamma_\mathrm{max}\approx 32$, and $L_{o,\mathrm{max}}\sim
10^{26}$\,W\,Hz$^{-1}$.  Figure~\ref{beta_lum} shows the distribution of
observed values in the $\beta_\mathrm{app}$ -- luminosity plane.  There
are no low luminosity sources with fast motions, but the high luminosity
sources show a wide range of apparent speeds.  The solid line represents
the locus of points corresponding to $\gamma_\mathrm{max}\approx 32$,
and $L_{o,\mathrm{max}}\sim 10^{26}$\,W\,Hz$^{-1}$ which forms a close
envelope to the data.  The curve has a peak value near the critical
angle $\theta_c=\gamma^{-1}\sim 3$~deg.   \cite{VC94} and \cite{LM97}
have shown that the most probable angle to observe a source is when
$\theta \sim 0.6\theta_c$ where $\beta_\mathrm{app}\sim 0.9\gamma$, so
in any jet sample there will be some jets which are oriented close to
$\theta_\mathrm{c}$ where $\beta_\mathrm{app}\approx\gamma$. \cite{C06}
have concluded from the observed speed --- luminosity distribution that
$\gamma_\mathrm{max}\sim 32$ and that intrinsic luminosities may extend
up to about $10^{26}$\,W\,Hz$^{-1}$.

\cite{C06} have called attention to the galaxies and BL Lacs located in
the lower left part of Figure~\ref{beta_lum}.  These sources are
unlikely to be powerful quasar jets oriented close to the plane of the
sky as might be expected from simple unified models \citep{UP95}, nor
can they be high -- $\gamma$, low -- $L_{o}$ jets beamed very close to
the line of sight as the probability of these configurations is too
low.  Instead, in general, the galaxies and BL~Lacs must form a separate
class of low luminosity jets.  However, the powerful radio galaxy
\object{Cygnus~A} has only a relatively weak jet with $0.59 < \beta <
0.68$. Following the simulations of \cite{A01}, \cite{C06} have
suggested that the \object{Cygnus~A} jet has a spine-sheath structure
with an energetic fast spine beamed away from the line of sight to feed
the powerful extended lobes and enshrouded by a slower observed sheath
whose broader but weaker beam encompasses the line of sight.

\subsection{The $\beta_\mathrm{app}$ -- $T_\mathrm{b}$ relation}
\label{sec:beta_T}

The brightness temperature of jet features may be determined directly
from observations, or estimated from the time scale of flux density
variations. Each of these methods gives separate insight to the physics of AGN
jets. The measured peak brightness temperature is typically in the range
$10^{11-13}$~K but extends up to $5\times 10^{13}$~K \citep{2cmPaperIV}. 
Observations with longer baselines, such as will be possible with the
planned Russian and Japanese space VLBI missions, will be needed to
determine whether higher brightness features can exist.

\begin{figure}[t!]
\begin{center}
\resizebox{\hsize}{!}{
   \includegraphics[trim = 9.8cm 13.2cm 0.9cm 0.6cm]{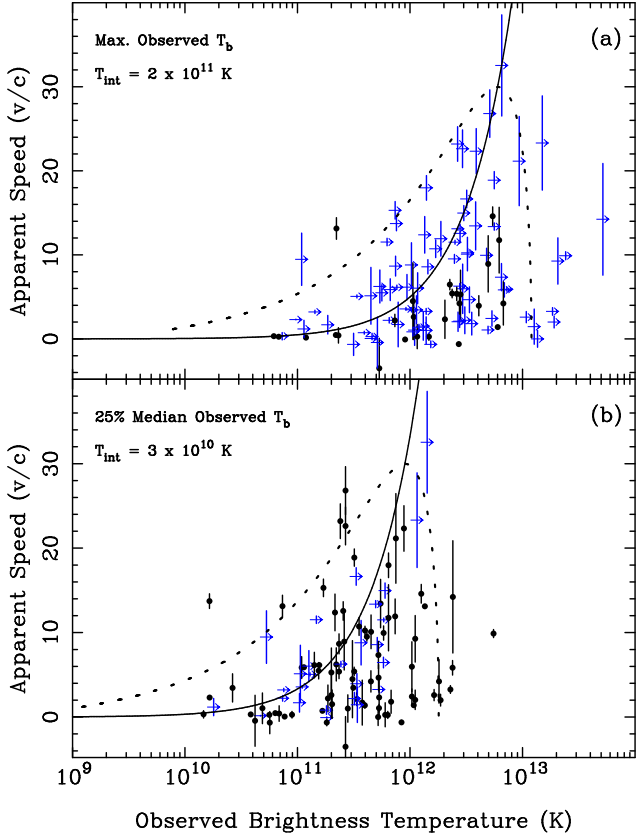}
}
\end{center}
\caption{\label{beta_T}
Plots of apparent speed, $\beta_\mathrm{app}$, vs.\ observed peak
jet brightness temperature, $T_\mathrm{b}$.  Lower limits are indicated
in blue by arrows.   The upper panel is based on the peak value of
$T_\mathrm{b}$ observed for each source, while the lower panel contains a
similar plot, except that it shows sources in their median-low state,
what \cite{H06} called the ``25\% Median.''  This 25\% median is the
median  of the lowest half of the brightness temperature observations
for a given source and represents a typical low brightness state for
each source. The solid line shown in each plot represents sources
observed at the  critical angle that has the intrinsic brightness
temperature indicated  in the upper left hand corner of the panel. The
dashed line represents the possible apparent speeds of a $\gamma = 30$
source with intrinsic brightness temperature given by the value of
$T_\mathrm{int}$. Adapted from \cite{H06}.
}
\end{figure}

\cite{H06} have used the peak brightness temperature at the base of each
jet observed by \cite{2cmPaperIV} to study the dependence of apparent
speed, $\beta_\mathrm{app}$, on apparent peak brightness temperature,
$T_\mathrm{b}$. Figure~\ref{beta_T} shows plots of $\beta_\mathrm{app}$
versus the maximum observed brightness (upper plot) temperature and
versus a more typical ``median-low'' value of brightness temperature. In
both cases, the solid line represents sources observed at the critical
angle ($\gamma^{-1}$) corresponding to the value of intrinsic brightness
temperature,$T_\mathrm{int}$, and the same value for $T_\mathrm{int}$ is
used in calculating the dotted ``envelope.''  These values of
$T_\mathrm{int}$ were chosen so that approximately 75\% of the sources
would fall below and to the right of the solid line corresponding to
what was found from simulations reported by \cite{H06}.

In both plots, the highest velocities are seen only for sources with
high apparent brightness temperature, and there is a clear absence of
fast sources with low brightness temperatures, as is also indicated by
the simulations.  \cite{H06} have therefore concluded that there is a
relatively narrow range of intrinsic brightness temperatures extending
perhaps a factor of two on either side of the solid line shown in Figure
\ref{beta_T}.  Since, the plot of maximum brightness temperature
contains more lower limits than measurements,  when in their high state,
the brightness temperature must be more than $2 \times 10^{11}$~K, which
corresponds to a large excess of particle energy over magnetic energy
\citep{KPT69,R94}. However, in the median-low state, $T_\mathrm{int}$ is
closer to the equilibrium value where the particle energy is about the
same as the magnetic energy.

\subsection{The $\beta_\mathrm{app}$ -- $D_\mathrm{var}$ relation}
\label{sec:beta_delta}

Most compact radio sources are variable on time scales typically of
months to years, \citep[e.g.,][]{KKNB02,AAH06}. An independent estimate
of the Doppler factor can be obtained from the time scale of the observed
variability. In the absence of any relativistic effects, the linear
dimensions are limited to the light travel distance on the time scale of
the variability. This puts a theoretical upper limit on the linear size
and consequently an upper limit on the angular size and corresponding
lower limit to the brightness temperature.  However, due to relativistic
boosting, the timescale for variability (and thus $T_{var}$) as well as
for $\beta_\mathrm{app}$ are both compressed. The apparent brightness
temperature in this case is given by  
\begin{equation}
T_{var} = \delta^{3}T_\mathrm{int}\,.
\label{eq:Tb_var}
\end{equation}

Equation \ref{eq:Tb_var} differs from equation \ref{eq:Tb} in that the
observed brightness temperature depends on $\delta^3$ rather than
$\delta$. In equation \ref{eq:Tb_var} two powers of $\delta$ are due to
the solid angle which is determined from the time variability. The third
factor of $\delta$ follows from the beaming as given in equation
\ref{eq:lum}.  

\begin{figure}[t]
\begin{center}
\resizebox{\hsize}{!}{
   \includegraphics[trim = 1.1cm 7cm 1.7cm 10.1cm]{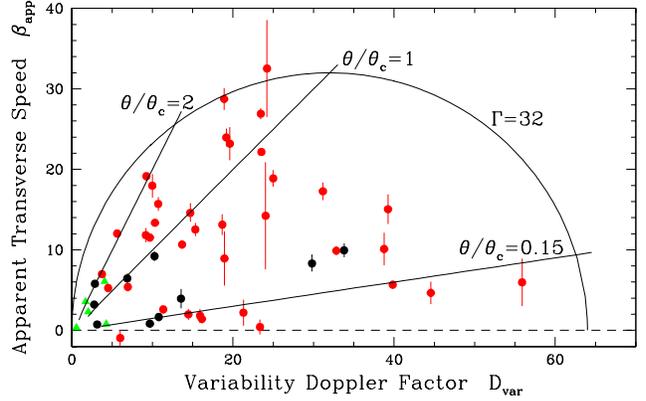}
}
\end{center}
\caption{\label{beta_D}
Plot of apparent transverse speed, $\beta_\mathrm{app}$, vs.\ 
$D_\mathrm{var}$ for components in 49 sources in the NRAO 2-cm VLBA
survey. The Doppler factor, $D_\mathrm{var}$ is derived from the
variability time scale near 1~cm, assuming an intrinsic brightness
temperature of $2\times 10^{10}$~K. The semi circle represents the locus
of points with a Lorentz factor, $\gamma  = 32$ as $\theta$
varies from 0 to 360 degrees. The
straight lines represent angles with respect to the line of sight in
terms of the critical angle  $\theta_\mathrm{c}$ where
$\delta_\mathrm{var} = \beta_\mathrm{app}$.  Red open circles are
quasars; blue full circles, BL~Lacs; and green triangles, galaxies.
Adopted from \cite{2cmPaperIII}.
}
\end{figure}

Following \cite{LH99}, \cite{2cmPaperIII} and \cite{C03} estimated the
Doppler factor, $\delta_\mathrm{var}$, for 49 sources from the time
scale of flux density variability measured near 1~cm. The plot for
$T_\mathrm{int} = 2 \times 10^{10}$~K gives the best fit between our
observed data and simulations based on a power law distribution of
luminosity and $\gamma$. Figure \ref{beta_D} shows a modified version
of $\beta_\mathrm{app}$, vs.\ $\delta_\mathrm{var}$ plot using updated
values of $\beta_\mathrm{app}$ from the MOJAVE program.  The best fit
value of $T_\mathrm{int} = 2 \times 10^{10}$~K is in good agreement with
the independently determined equilibrium value of $T_\mathrm{int}$ found
from the distribution of peak values of $T_\mathrm{b}$ obtained from
the direct VLBA measurements (Figure~\ref{beta_T}) and is significantly
less than the inverse Compton limit of $\sim 5 \times 10^{11}$~K.  

\subsection{Finite opening angles}
\label{sec:opening}

The quantitative discussion in this section assumes that the jet is
one dimensional.  \cite{GWD06} and \cite{GDSW06} have discussed the
implications of finite opening angles and a non uniform velocity across
the jet cone. They also consider the special case where the viewer is
inside the opening angle of the jet.  They argue that if the jet opening
angle is sufficiently large, then the Lorentz factors calculated under
the assumption of a one dimensional jet may be substantially
underestimated.  

\section{Radio Jets and Gamma-Ray Emission}
\label{sec:gamma}

Many gamma-ray bright sources are identified with flat spectrum
radio AGN \citep{M01,H03,SRM03}. However, comparison of radio jet
structure or kinematics with gamma-ray emission has been inconclusive. 
\cite{2cmPaperI} reported no difference in the morphology of AGN jets
with and without observed gamma-ray emission, but \cite{2cmPaperIV}
found that the radio jets of gamma-ray sources appear to have more
compact structure than non gamma-ray sources. \cite{2cmPaperIII} and
\cite{J01} reported evidence that AGN with observed gamma-ray emission
have somewhat higher observed speeds, but this was based on the very
limited gamma-ray data obtained by the EGRET detector on the Gamma Ray
Observatory.  The start of GLAST operations in 2008 will give greatly
improved sensitivity and time sampling for thousands of gamma-ray
sources and will establish whether there is a substantial increase in
the number of gamma-ray detections, or if instead, there is a class of
``gamma-ray quiet'' quasars analogous to radio-quiet quasars. Our
planned extension of the MOJAVE program through the GLAST era will give
us a better understanding of the relation between the radio jets and
gamma-ray emission.

\section{Summary}
\label{sec:summary}

VLBA observations have led to an increased understanding of the nature
of AGN jets including the following:
\\
1. Trends of observed velocity with luminosity, brightness temperature, and
variability are consistent with relativistic beaming models in which
blazar jets are highly relativistic with Lorentz factors extending up to
about 30 and are oriented close the line of sight.
\\
2. There is a broad distribution of Lorentz factors among observed
sources.  The parent population contains jets which are mostly only
mildly relativistic.  However, due to Doppler boosting bias, in flux
density limited samples we see mostly the ultra-relativistic jets.  But,
some of the slower jets are also directly observed.
\\
3. The collimation and acceleration appear to occur close to the
central engine, but {\it in situ\/} acceleration within the jet is also
important, and the final collimation in some cases may be determined up to a
kiloparsec or more downstream.
\\
4. Each jet appears to have a characteristic velocity which may
reflect the bulk plasma flow.  In general, features observed at shorter
wavelengths, which are closer to the core, appear to move with higher
speed.  Quasar jets typically have apparent speeds around
$10c$, corresponding to intrinsic speeds of 99.9~percent of the speed of
light. Apparent speeds up to about $35c$ are observed in some quasars.  
Lower luminosity AGN, the cores of extended FR~I and FR~II radio
galaxies, and BL~Lac objects usually appear much slower with apparent
velocities which are close to the speed of light and may even appear
subluminal.  
\\
5. Radio jets associated with gamma-ray sources appear to have, on average,
more compact structure and show faster outflow speeds.
\\
6. Initially, following the injection of new relativistic particles,
intrinsic (in the source frame) brightness temperatures up to a few times
$T_\mathrm{int} \sim 10^{11}$ K are inferred corresponding to an apparent excess
of particle energy over the energy in magnetic fields. But
the energy balance later approaches equilibrium corresponding to
$T_\mathrm{int} \sim 5 \times 10^{10}$~K.
\\
7. The overall kinematics of the jets are rather complex, with some jets
displaying bright features moving on linear trajectories, while others
have accelerated motions on curved trajectories. In many sources,
successive features are ejected on different sky position angles, and do
not follow the paths of previous knots. 
\\
8. The quantitative analysis and derivation of intrinsic properties may be
complicated by possible differences between the pattern and bulk
velocity flow as well as by the finite opening angle of jets and
possible velocity or density gradients across the jet.

In the next few years, improved data recording systems will allow better
sensitivity so that it will be possible to trace the structure further
along the jets and to follow the motion of individual features for a
longer time. The improved sensitivity will also permit more routine
observations at shorter wavelengths with corresponding better angular
resolution. The availability of the GLAST all-sky gamma-ray data with 
high sensitivity and good time 
resolution should enhance our understanding of the connections 
between radio jets and gamma-ray emission.

\section{Data Archive}

All of our images, movies, the RATAN
spectral monitoring from 1.4 to 30~cm, and all kinematic
data are available on our web
site\footnote{http://www.physics.purdue.edu/astro/MOJAVE/}. Flux
densities measured at the UMRAO at 2, 4, and 6 cm can be found at the
UMRAO
Database\footnote{http://www.astro.lsa.umich.edu/obs/radiotel/umrao.html}
and more up to date data may be obtained by contacting the UMRAO
authors.

Since 2002, we have extended our observations to include measurements of
circular and linear polarization \citep{LH05,HL06}. Polarization data
are  given on our web site. However, discussion of the polarization
properties of AGN jets is beyond the scope of this paper. 

\begin{acknowledgments}
This paper is based on observations made with the  Very Long Baseline
Array which is a facility of the National Radio Astronomy Observatory
which is operated by Associated Universities, Inc., under a cooperative
agreement with the National Science Foundation. Part of this work was
done by YYK, MLL and DCH during their Karl Jansky postdoctoral
fellowship at the National Radio Astronomy Observatory. YYK is currently
a Research Fellow of the Alexander von Humboldt Foundation.   DCH was
partially  supported by an award from the Research Corporation.  MK was
supported  in part through a stipend from the International Max Planck
Research School for Radio Astronomy at the University of Bonn and, in
part, by a NASA Postdoctoral Program Fellowship appointment at the
Goddard Space Flight Center.  The MOJAVE project is supported under
National Science Foundation grant AST-0406923 and a grant from the
Purdue Research Foundation.  The work at the \mbox{RATAN-600}  radio
telescope was supported by the  Russian Ministry of Education and
Science, the NASA JURRISS Program  (project W-19611), and the Russian
Foundation for Basic Research  (grants 01-02-16812 and 05-02-17377). The
University of Michigan Radio Astronomy Observatory is supported by the
University of Michigan and the National Science Foundation (grant
AST-0607523).  
\end{acknowledgments}

\bibliographystyle{spr-mp-nameyear}

\bibliography{yyk}

\end{document}